\documentclass[vecphys]{svmult}

\usepackage{makeidx}         
\usepackage{graphicx}        
\usepackage{natbib}        
\usepackage{multicol}        
\usepackage[bottom]{footmisc}

\newcommand\aj{{AJ}}
\newcommand\araa{{ARA\&A}}
\newcommand\apj{{ApJ}}
\newcommand\apjs{{ApJS}}
\newcommand\aap{{A\&A}}
\newcommand\aaps{{A\&AS}}
\newcommand\mnras{{MNRAS}}

\makeindex             

\begin{document}

\title*{Population synthesis at short wavelengths and spectrophotometric 
diagnostic tools for galaxy evolution}
\titlerunning{Population synthesis at short wavelengths}

\author{Alberto Buzzoni\inst{1}\and
Emanuele Bertone\inst{2}\and
Miguel Ch\'avez\inst{2}\inst{,3}\and
Lino H. Rodr\'\i guez-Merino\inst{2}
}

\authorrunning{Buzzoni et al.}
\institute{INAF - Osservatorio Astronomico di Bologna, Via Ranzani 1 - 40127 Bologna, Italy - 
\texttt{(e-mail: alberto.buzzoni@oabo.inaf.it)}
\and INAOE - Instituto Nacional de Astrof\'\i sica, Optica y Electr\'onica,
Luis Enrique Erro 1 - 72840 Tonantzintla, Pue, Mexico -
\texttt{e-mail: (ebertone@inaoep.mx}, \texttt{mchavez@inaoep.mx}, \texttt{lino@inaoep.mx)}
\and
IAM - Instituto de Astronomia y Meteorologia, Universidad de Guadalajara, Vallarta 2602 - 
44130 Guadalajara, Mexico
}
\maketitle

\begin{abstract}
Taking advantage of recent important advances in the calculation of high-resolution
spectral grids of stellar atmospheres at short-wavelengths, and their implementation 
for population synthesis models, we briefly review here some special properties of 
ultraviolet emission in SSPs, and discuss their potential applications for identifying and 
tuning up effective diagnostic tools to probe distinctive evolutionary properties of 
early-type galaxies and other evolved stellar systems.
\end{abstract}

\section{Introduction}

With an amazingly successful series of dedicated space missions, the pioneering 70's marked 
the beginning of ultraviolet astronomy; satellites like {\sc ANS}, {\sc OAO}, and {\sc IUE} 
opened the way, in fact, to the exploration of the nearby Universe at short spectral wavelength 
by-passing, for the first time, the blocking effect of Earth atmosphere.
Since then, ultraviolet astronomy has received a renewed impulse in the current decade,
partly due to on-going space projects like the {\sc GALEX} mission or even the {\sc HST}, 
but also under much different observational circumstances, as more powerful ground-based optical
telescopes made the redshifted short-wavelength emission of distant galaxies to be 
eventually detectable at their visual-infrared eyes.

Curiously enough, we ended up by browsing ultraviolet features of galaxies at cosmic distances at much 
finer detail\footnote{In addition, redshift acts on spectroscopic observations by improving 
wavelength resolution $\lambda/\Delta\lambda$ by a factor $(1+z)$ in the galaxy restframe.} 
compared to a still relatively scanty survey of the local stellar systems 
at wavelengths outside the optical range.
For this reason, our understanding of the deep Universe cannot fully rely on a straightforward
application of the local empirical templates, but rather needs an added value 
by theory to assess the evolution of high-redshift of stellar populations.

In this regard, population synthesis has attained nowadays an unprecedented accuracy in reproducing 
galaxy spectral features thanks to the match with increasingly refined libraries of stellar 
model atmospheres and improved algorithms to derive therefrom the 
high-resolution spectral information across the widest range of fundamental parameters 
(i.e. $\log T_{\rm eff}, \log g,$ and [Fe/H]). As far as the short-wavelength range is 
concerned (i.e. for $\lambda$ $_<\atop^{\sim}$ 3000~\AA)  we can presently count on 
important theoretical datasets covering the full parameter space to reproduce 
stellar spectral energy distribution (SED) across the whole H-R diagram, at wavelength 
resolution typically better than $\lambda/\Delta\lambda$ $_>\atop^{\sim}$ $20\,000$ 
\citep[see][for an exhaustive review on this subject]{bertone05}.
Among others, these include the work of \citet{lino} ({\sc Uvblue} spectral library), 
\citet{gustaffson03} ({\sc Marcs} model grid), \citet{munari}, \citet{hauschildt99} 
(the {\sc NextGen} library), and \citet{coelho}.

As an effort to settle the interpretative framework that stems from the analysis of the 
short-wavelength SED of stellar systems, in this contribution we want to briefly review some 
special properties of ultraviolet emission in simple stellar populations (SSPs), and 
discuss their potential applications for identifying and tuning up effective diagnostic tools 
for population synthesis studies.

\section{Short- vs. Long-term memory}

It is commonly recognized that UV luminosity carries direct information on the age of young SSPs,
being sensitive to the presence of hot and bright high-mass stars at the top main sequence 
(MS) turn-off (TO) point \citep[e.g.][]{oconnell99}. While this certainly holds for a starburst case,
where colors like $(U-V)$ are actually fair age tracers \citep[see, e.g.][]{leitherer99,mashesse91},
things can be different, and subtly more entangled, in case star formation (SF) proceeds allover
the galaxy life.

The key feature in this framework resides in a physically different ``regime'' of UV luminosity 
evolution compared to longer wavelengths. 
As far as the Johnson $U$ band is concerned, for instance, models indicate that, for a solar metallicity
and a Salpeter slope for the power-law IMF, SSP luminosity scales with time as 
\begin{equation}
L_{\rm SSP}^{(U)}\quad \propto \quad t^{-1.1} 
\label{eq:1}
\end{equation}
\citep[see, e.g.][]{buzzoni05}. This holds over virtually the whole SSP history, from ages of a few Myr up 
to 10 Gyr and beyond (see Fig.~1).

\begin{figure}[t]
\begin{minipage}{0.56\hsize}
\includegraphics[width=\hsize]{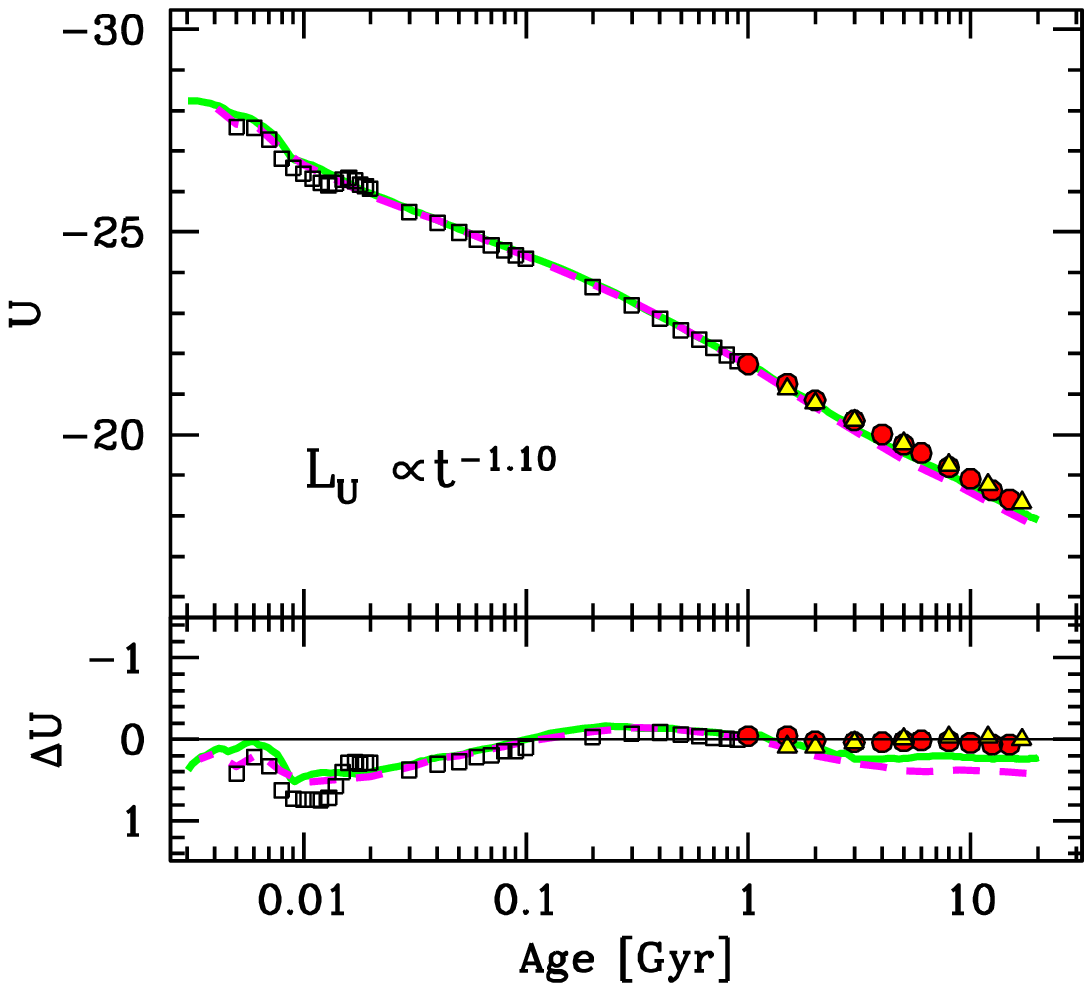}
\end{minipage}
\begin{minipage}{0.43\hsize}
\caption{Theoretical SSP luminosity evolution for solar metallicity and Salpeter IMF, according to different
population synthesis codes, namely \citet{buzzoni89,buzzoni05} (dots), \citet{worthey} (triangles), \citet{leitherer99}
(squares), \citet{bressan94} (dashed line), and \citet{bc03} (solid line). Absolute $U$ magnitudes are
scaled to a total SSP mass of $10^{11}$~M$_\odot$. For each model sequence, the residuals with respect to a 
perfect $L_U \propto t^{-1.10}$ trend are shown in the lower panel.}
\end{minipage}
\label{fig:1}
\end{figure}

If we assemble several equally mass-weighted SSPs (each producing stars up to a mass M$_{\rm up}$ with a fixed 
IMF) along a wider age range such as to smoothly reproduce a constant SF rate over time $t$, then the total 
luminosity of the resulting composite stellar population (CSP) becomes 

\begin{equation}
{\cal L}_{\rm gal}(t) \propto \int^t_{t_o} L_{\rm SSP}(\tau) d\tau = \int^t_{t_o} \tau^{-\alpha} d\tau =
{1\over {1-\alpha}}\quad \left[ t^{1-\alpha} - t_o^{1-\alpha}\right].
\label{eq:2}
\end{equation}

In the equation, $t_o$ is the lifetime of stars of highest mass, M$_{\rm up}$, and 
can be operationally (and physically) conceived as the discrete integration time step $d\tau$ in each
summation.
If $\alpha < 1$, one sees that the r.h. solution of the integral is actually
modulated by the {\it oldest} SSP components (that is those stars about $t$ years old)
as, in general,  $ t^{1-\alpha} \gg t_o^{1-\alpha}$. On the contrary, if the SSP luminosity fades
more rapidly than $L_{\rm SSP} \propto t^{-1}$, that is for $\alpha > 1$, then the term 
$t_o^{1-\alpha}$ prevails and most of CSP luminosity comes
from the {\it youngest} composing SSPs. In the latter case, 
\begin{equation}
{\cal L}_{\rm gal} \propto {{t_o^{1-\alpha}} \over {\alpha -1}} = {\rm const.(M_{up},\alpha)}
\label{eq:3}
\end{equation}
and the integrated CSP luminosity looses any dependence on age, only responding to the amount
of short-lived high-mass stars, as modulated by the actual SF rate of the stellar aggregate 
\citep[see][for a more detailed discussion]{buzzoni02}.
For the case of the $U$ band $\alpha = 1.1$, and this is why ultraviolet luminosity 
is a so effective tracer of the {\it actual} SF activity carrying therefore ``short-term'' memory 
of the CSP history.
More generally, $\alpha$ is expected to vary as a function of wavelength, as well, as the slope
in the $\log L$ vs. $\log t$ relationship for a SSP depends on the relative contribution
to total luminosity of stars in the different regions of the H-R diagram.

\begin{figure}[t]
\centering
\includegraphics[width=0.84\hsize]{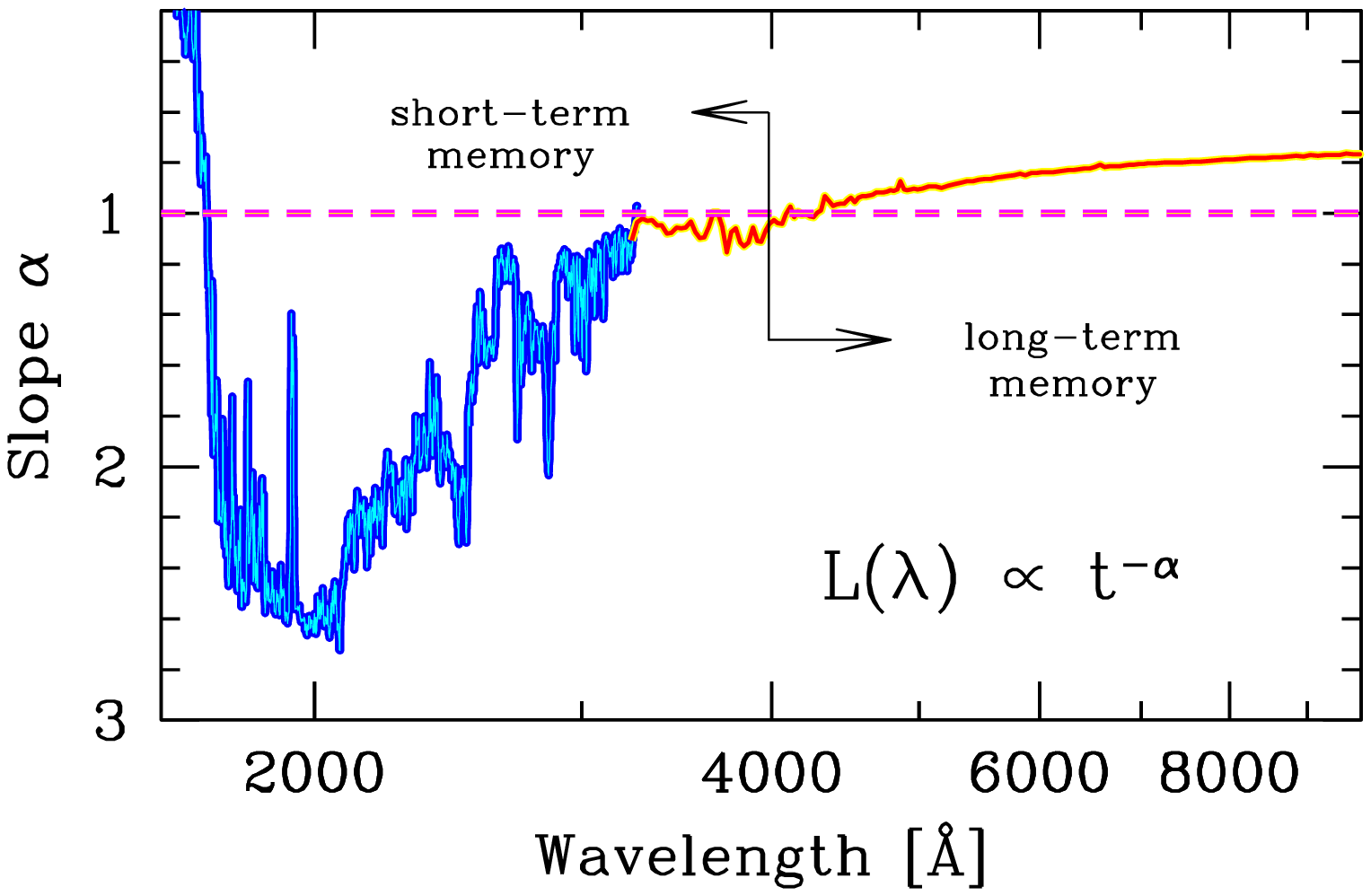}
\caption{Expected change with wavelength of the slope $\alpha = -\partial \log L(\lambda) /\partial \log t$ 
for a 15 Gyr SSP with solar metallicity, Salpeter IMF, and a red horizontal branch morphology, 
according to \citet{buzzoni89}. High-resolution UV synthesis made use of the {\sc Uvblue} 
spectral library of \citet{lino}. Note that $\alpha$ exceeds unity in the ultraviolet 
spectral region shortward of 4000~\AA. As discussed in the text, this threshold value discriminates between 
two different regimes for SSP luminosity evolution.
}
\label{fig:2}
\end{figure}

For the same reference SSP of Fig.~1, this is shown in Fig.~2 along the full 
spectral range.
From the figure, one notes for example that eq.~(\ref{eq:3}) holds along the entire mid-UV wavelength 
range, being $\alpha > 1$ for $\lambda$ $_>\atop^{\sim}$ 4000 \AA. On the contrary, the visual and infrared 
ranges characterize for a shallower SSP luminosity evolution ($\alpha < 1$). When applied to the case of a
CSP, therefore, colors like $V-K$ are expected to carry ``long-term'' memory of CSP history through the 
cumulative effect of long-lived unevolved stars of low mass along the entire life of the stellar system.

The transition between short- and long-term memory regimes along SED of a CSP will also depend on the IMF details.
Compared to the Salpeter case, for example, a stellar aggregate displaying a steeper 
(i.e. giant-dominated) IMF power slope would more likely wipe out any sign of its more remote past  
being populated at any time, on average, by shorter-lived higher-mass stars.
In this case, $\alpha$ will exceed unity well longward of $\lambda \simeq 4000$~\AA.

\section{The Age-metallicity degeneracy}

In spite of any more or less exotic recipe to combine stellar tracks and isochrones, 
from the physical point of view, the ultimate driving parameter that eventually constrains SED
of a SSP is the mass of TO stars (M$_{\rm TO}$). The latter will in fact set the cosmic clock 
(through MS stellar lifetimes) and reverberate on the overall morphology of the different
evolutionary stages across the synthetic H-R diagram of the SSP. As a major drawback of this situation, 
anytime we try to derive an absolute age estimate for the (either resolved or unresolved) SSP we also need to 
set {\it at the same time} its chemical composition. In other words, age
and metallicity are intimately tied such as a wide range of SSPs can in principle give rise
to fully equivalent spectrohotometric outputs. 

More generally, synthesis models have extensively demonstrated that {\it ``a factor of three change in age
produces the same change in most colors and indices as a factor of two in $Z$''} \citep{worthey92}.
So, old metal poor SSPs closely resemble in color and overall SED younger metal-rich
ones. This effect, known as the ``age-metallicity degeneracy'' \citep{rb86,buzzoni95}, cannot easily be 
overcome as far as we restrain our analysis to the optical range of the SED of stellar systems, and we 
are forced therefore to browse the most extreme spectral windows, both at shorter and longer wavelength
range, to break the ``3-to-2'' degeneracy and decouple, in principle, the $t$ and $Z$ pieces of information 
in SSPs. In this regard, ultraviolet is certainly a preferred window, in force of the more selective
dependence of SSP luminosity on the upper MS stellar component.

Again, taking the SED of a 15 Gyr Salpeter SSP of solar metallicity as a reference, we have estimated 
in Fig.~3 the expected changes of total monochromatic luminosity along the entire SED  vs.\ a change 
either in metallicity or age, assuming that

\begin{equation}
\left\{
\begin{array}{l}
L(15,Z) \propto L(15,Z_\odot) \left({Z\over Z_\odot}\right)^{-\beta(\lambda)} ~\qquad\qquad {\rm for~fixed~age,~or}\\
L(t,Z_\odot) \propto L(15,Z_\odot) \left({t\over {15{\rm Gyr}}}\right)^{-\alpha(\lambda)} \qquad\qquad {\rm for~fixed~metallicity}.
\end{array}
\label{eq:4}
\right.
\end{equation}

Worthey's steep slope $\Delta \beta/\Delta \alpha = \partial \log t /\partial \log Z \simeq \log 3/\log 2$ 
can easily be recognized for a long wavelength path spanning from the near infrared to the optical
window, but a more complex behaviour begins to appear in the ultraviolet, especially in the Mid-UV range
around the spectral region that roughly corresponds to the {\sc Galex} NUV pass-band, as displayed 
on the plot. The $\beta$ vs.\ $\alpha$ curve then turns about 2000 \AA\ changing direction 
with decreasing $\lambda$ and eventually flattening shortward of 1500 \AA.\footnote{Note that, contrary to the
visual wavelenght range, age-metallicity degeneracy shortward of 2000 \AA\ behaves in the opposite way, 
as the effect on FUV colors of a {\it younger} SSP can be recovered by {\it decreasing} metallicity.}

\begin{figure}[t]
\centering
\includegraphics[width=0.84\hsize]{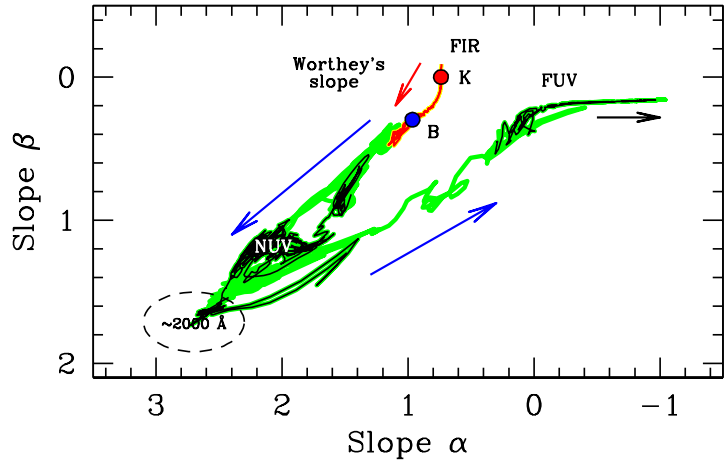}
\caption{Theoretical estimate of $\alpha = -\partial \log L(\lambda)/\partial \log t$ and 
$\beta = -\partial \log L(\lambda)/\partial \log Z$ slopes with varying wavelength from the 
infrared (Johnson $K$ band at 2.2~$\mu m$) to the far ultraviolet ($\lambda \sim 1500$~\AA),
as indicated by the arrows,  for the reference SSP of Fig.~2. Johnson $B$ and $K$ wave bands are labelled 
for reference on the plot, as well as
the {\sc Galex} Far- (FUV) and Near-UV (NUV) bands. Note the turn-around feature about 2000~\AA,
and an opposite trend of the $\beta$ vs.\ $\alpha$ relationship when moving toward the Far-UV spectral range.
The \citet{worthey92} ``3-to-2'' degeneracy vector is displayed on the plot (see text for a discussion).
}
\label{fig:3}
\end{figure}
 
\section{UV indices and SSP diagnostic}

The recent theoretical grids of synthetic stellar atmospheres, covering the short-wavelength range
at high spectral resolution, have made possible to directly trace the evolution of specific spectral features
along the ultraviolet region of the SED of synthetic stellar atmospheres.

In particular, the mid-UV (i.e. 2200 $_<\atop^{\sim}$ $\lambda$ $_<\atop^{\sim}$\ 3200 \AA) wavelength range 
has been carefully screened by \citet{chavez07}, based on the \citet{lino} {\sc Uvblue} spectral grid.
The Chavez et al.\ analysis tackles, from the theoretical side, the original work
of \citet{fanelli90}, who carried out a comprehensive analysis of IUE stellar database through a set of
narrow-band spectrophotometric indices able, in principle, to independently probe temperature, surface 
gravity, and chemical composition of stars. The  big advantage of the theoretical approach, over the 
empirical one, however, is that one can more comfortably explore the change of spectral 
features with varying (in a controlled way) either one or more of the stellar fundamental parameters. 
In addition, also obvious limits of empirical samples can be overcome, like for instance the narrow range 
of [Fe/H] distribution, naturally peaked around the solar value when observing real stars in the solar 
neighbourhood.

The \citet{chavez07} theoretical framework has been further extended, by matching the {\sc Uvblue}
spectral library with the \citet{buzzoni89} population synthesis code to obtain synthetic 
UV indices for SSPs. Full details of this project are the subject of a forthcoming paper
\citep{bertone07}, but we want to assess here just a few important issues related to a more refined use
of narrow-band indices {\it \'a la Fanelli} to probe evolutionary parameters of stellar populations. 
According to our previous discussion, as far as SSPs are concerned, the mid-UV spectral region appears to 
be the most promising one to break the ``3-to-2'' age-metallicity degeneracy. On a narrow-band wavelength
scale like in Fig.~4, in fact, one can appreciate that the $\beta$ vs. $\alpha$ correlation 
drastically deviates from the Worthey's slope, providing any sort of spectral ``leverage'' suitable
to provide decoupled or even {\it orthogonal} pieces of information about age and metallicity. In this 
regard, by properly chosing both feature and pseudo-continuum bands, such as to have either a prevailing 
dependence on $\beta$ or $\alpha$, one could set up, in principle, new and optimzed narrow-band indices 
either metal- or age-sensitive, respectively.

\begin{figure}[t]
\begin{minipage}{0.62\hsize}
\includegraphics[width=\hsize]{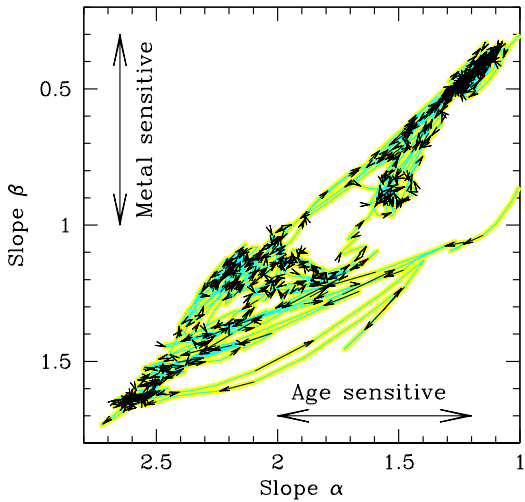}
\end{minipage}
\begin{minipage}{0.36\hsize}
\caption{A zoomed plot of the UV region from  Fig.~3, sketching the trend of the 
$\Delta \beta/\Delta \alpha = \partial \log t /\partial \log Z$ degeneracy vector along
wavelength, from 3000~\AA\ (top right corner) to $\sim 1500$~\AA\ (mid-range curve to the right).
According to the slope of the  $\beta$ vs.\ $\alpha$ relation, the different intervals of the UV SED
can be selectively sensitive to either age ($\Delta \beta \to 0$) or metallicity changes
($\Delta \alpha \to 0$).}
\end{minipage}
\label{fig:4}
\end{figure}

On the same line, by means of Fig.~4 we could easily probe age- and metal-sensitivity of 
established narrow-band indices, like those in the \citet{fanelli90} system as well.
Two interesting examples of nearly ``horizontal'' (i.e. most age-sensitive) and ``vertical'' 
(i.e. most metal-sensitive) indices are displayed in Fig.~5, for the case of the 
2332~\AA\ Fe~{\sc ii} feature and for the 2538~\AA\ metal blend.

Given its manifold piece of information, we are going to more systematically 
explore $\beta$ vs.\ $\alpha$ diagnostic diagrams like those of Fig.~4 in order to 
identify and exploit potentially useful features to more cleanly probe distinctive properties of 
stellar population relying on the spectroscopic analysis of their integrated UV spectrum.
To a more refined approach, however, such an excercise needs to more accurately size up the influence
on Fig.~4 of theoretical uncertainties in high-resolution UV spectral synthesis, as well as the impact of
other distinctive parameters of stellar aggregates, like the horizontal branch 
morphology, $\alpha$-enhanced chemical partitions, and different slopes in the power-law IMF.
\vfill\eject

\begin{figure}[]
\centering
\includegraphics[width=0.48\hsize]{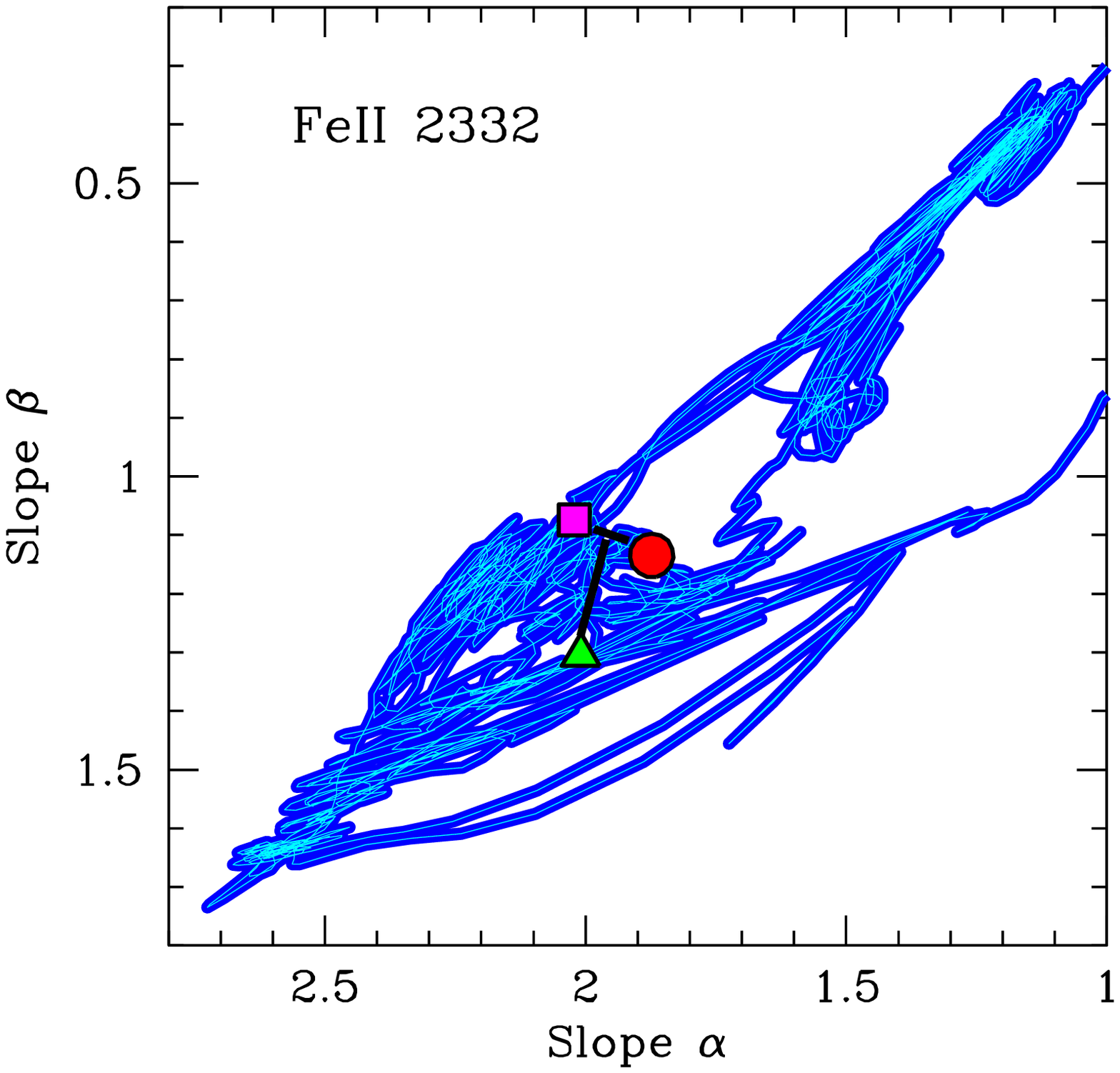}
\includegraphics[width=0.48\hsize]{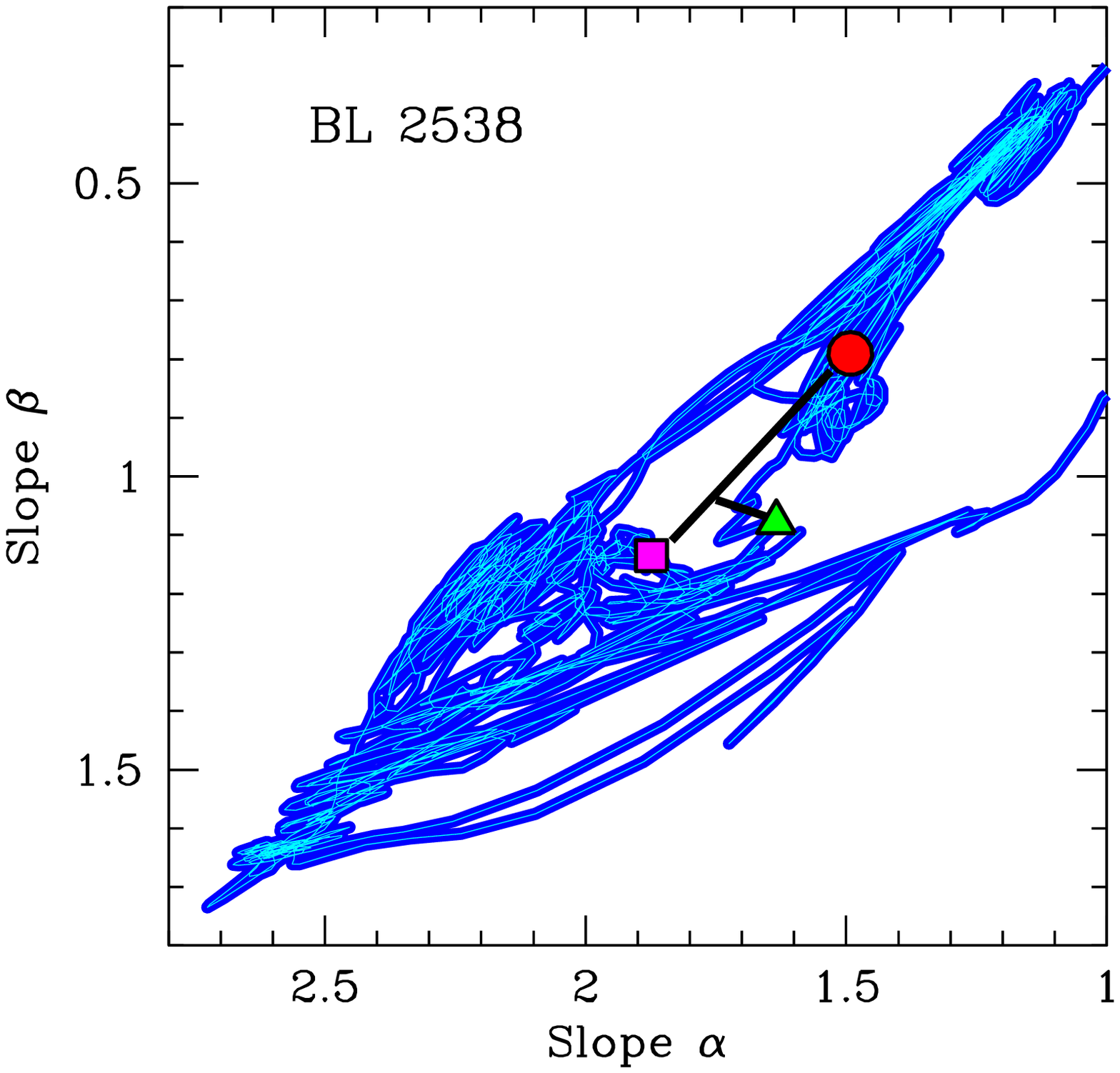}
\caption{Two illustrative cases of the \citet{fanelli90} narrow-band spectrophotometric indices located 
in the plot domain of Fig.~4. For the Fe~{\sc ii} feature at 2332~\AA\ (left panel) and the metal blend 
about 2538~\AA\  (right panel), the local degeneracy vector ($\Delta \beta/\Delta \alpha$) can be estimated 
by connecting the feature location (triangles on the plots) 
with the corresponding pseudocontinuum, as interpolated from two ``blue'' (squares) and ``red'' (dots) 
side bands. Note that, for its almost vertical slope, the Fe{\sc ii} index is better sensitive to 
metallicity, while thanks to a nearly horizontal trend, the Bl~2538 index will better respond to any 
change in SSP age.
}
\label{fig:5} 
\end{figure}

\noindent {\bf Acknowledgments -} We are pleased to acknowledge partial financial support for this project 
from Italian MIUR, under grant INAF PRIN/05 1.06.08.03, and from Mexican CONACyT, via grants 36547-E and 
SEP-2004-C01-47904.
\begin{scriptsize}
%

\end{scriptsize}
\printindex
\end{document}